\documentclass[aps,nofootinbib,twocolumn,superscriptaddress,letterpaper,amsmath,amssymb]{revtex4}

\usepackage{graphicx}
\usepackage{dcolumn}
\usepackage{bm}
\usepackage{amssymb}
\usepackage{feynmf}
\usepackage{slashed}
\usepackage{multirow}
\usepackage{color}
\usepackage{array}
\usepackage{url}

\usepackage{tikz}
\usepackage{siunitx}
\usepackage[breaklinks]{hyperref}
\usepackage[percent]{overpic}

\begin{document}

\title{Learning Broken Symmetries with Resimulation and Encouraged Invariance}

\author{Edmund Witkowski}
\affiliation{Department of Physics and Astronomy, University\ of\
  California, Irvine, CA\ 92697}
\author{Daniel Whiteson}
\affiliation{Department of Physics and Astronomy, University\ of\
  California, Irvine, CA\ 92697}

\begin{abstract}
Recognizing symmetries in data allows for significant boosts in neural network training.  In many cases, however, the underlying symmetry is present only in an idealized dataset, and is broken in the training data, due to effects such as arbitrary and/or non-uniform detector bin edges. Standard approaches, such as data augmentation or equivariant networks fail to represent the nature of the full, broken symmetry. We introduce a novel data-augmentation scheme that respects the true underlying symmetry and avoids artifacts by augmenting the training set with transformed pre-detector examples whose detector response is then resimulated. In addition, we encourage the network to treat the augmented copies identically, allowing it to learn the broken symmetry.
While the technique can be extended to other symmetries, we demonstrate its application on rotational symmetry in particle physics calorimeter images. We find that  neural networks trained with pre-detector rotations 
converge to a solution more quickly than networks trained with standard post-detector augmentation, and that networks modified to encourage similar internal treatment of augmentations of the same input converge even faster.
\end{abstract}

\date{\today}

\maketitle

\section{Introduction}
\label{sec:introduction}

Evidence for new physics and subtle features of the Standard Model are often hidden in high-volume, high-dimensional datasets produced at the Large Hadron Collider and in other high-intensity particle beams.  Traditional methods of data analysis reduce the dimensionality of the data with engineered features which exploit our physical understanding of the task.  While powerful, these heuristics often rely on simplifying assumptions which fail to fully capture the available information.  Recently, artificial neural networks have demonstrated the capacity to exceed the performance of engineered features~\cite{Baldi:2014kfa,deOliveira:2015xxd,Feickert:2021ajf}.
However, training such networks often requires vast quantities of data or computational resources, which can be problematic in practice\cite{Alzubaidi:2021review}. 
There may only be a limited amount of data available for a given region of interest, or there may be computational limitations on how much data can feasibly be processed or generated with simulation programs.  Learning strategies which are more efficient, reaching the performance plateau with fewer learning cycles or on smaller training samples, are therefore of great value to the particle physics research program.

 Efficiency may be gained by leveraging physical symmetries present in the data, where the data are closed under some transformation.
This is typically done by enforcing equivariance in latent space operations or by requiring invariance in classification output.\cite{Cohen:pmlr-v48-cohenc16,Agrawal:7410370,Gens:NIPS2014_f9be311e,Bogatskiy:2022czk,Shimmin:2023cfw,Shimmin:2021pkm,Bogatskiy:2022hub}
If the symmetry group is understood exactly, the network structure might incorporate it, effectively constraining the functional search space. A constrasting strategy is data augmentation, expanding the training set to explicitly include transformations of the original data, allowing the network to infer the symmetry. However, in many cases the symmetry is exact only in an idealized scenario and in practice is {\it broken} by asymmetries such as the data-collection devices.  For example, the exact and continuous translational and rotational symmetry of an idealized image of a cat is broken into a discrete symmetry by the camera's pixel edges. Arbitrary shifts or rotations of the cat only generate shifted or rotated versions of the image when they reflect the discrete symmetry of the pixel geometry.   If the pixels themselves are not identical, the symmetry is broken further. What was a powerful, continuous and exact symmetry is broken into a less effective, discrete and approximate symmetry, which hinders our ability to exploit it to boost training efficiency.

\begin{figure}[h!]
    \centering
    \includegraphics[width=0.95\linewidth]{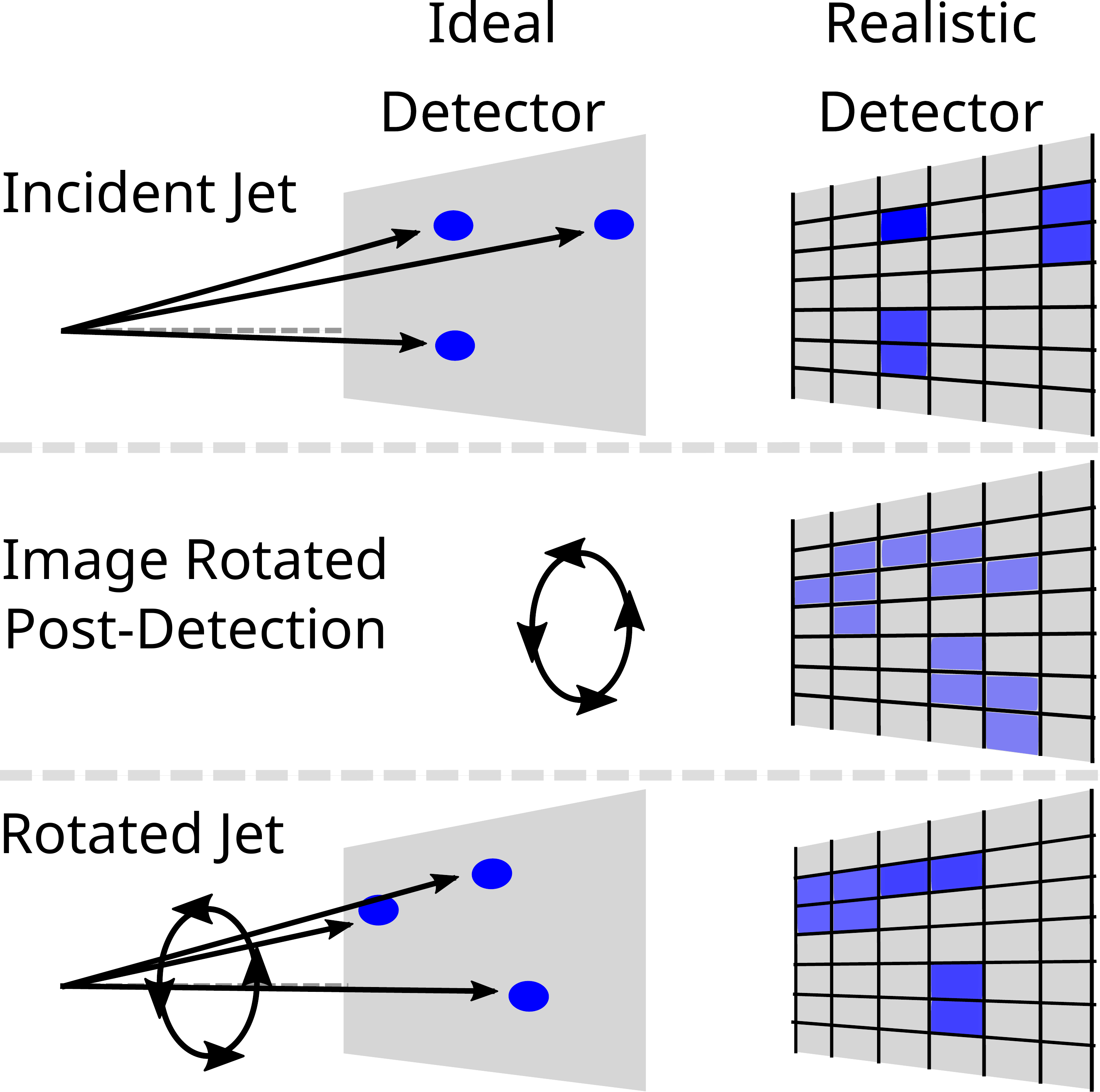}
    \caption{ Demonstration of the breaking of a continuous rotational symmetry by the pixelization of a realistic detector. (Left) shows an ideal detector that performs no binning, while (Right) is a realistic detector which produces a pixelated image. (Top) shows a jet incident on the detector, and the images produced by each. (Middle) shows the image produced by the realistic detector rotated by an angle that is not a multiple of $\frac{\pi}{2}$. Rotating a pixelated image by such an angle results in artifacts and does not produce a detector image which reflects the true symmetry. (Bottom) shows the case where the jet itself is rotated pre-detector, producing an image which accurately represents the symmetry of the problem. Though it is not closed under rotation, it avoids introducing artifacts from post-detector rotation.
    }
    \label{fig:binning_graphic}
\end{figure}

Detection of particle energies by detectors presents a widespread and important example of broken symmetries, especially in the context of detection of a jet's energy deposition by a grid of calorimeter cells.  A given jet is equally likely to have any orientation around its axis, and rotation of its constituent particles around the axis changes none of its crucial physical observables.  An idealized detector would respect this symmetry, but a realistic calorimeter is composed of discrete, non-uniform cells. The continuous symmetry is broken such that the detector pattern is preserved only under rotations of the particles by multiples of $\frac{\pi}{2}$.  Rotations of the observed image by arbitrary angles, as is often done in data-augmentation strategies, introduce artifacts from the double-pixelization and fail to generate training examples which demonstrate the true symmetry, as demonstrated in Fig.~\ref{fig:binning_graphic}.  Enforcing symmetry in the network's latent operations faces a similar hurdle, as the training data do not reflect the true continuous symmetry, only the more limited, broken symmetry.

\begin{figure}
    \centering
    \begin{minipage}{0.48\linewidth}
        \centering
        \includegraphics[width=0.95\linewidth]{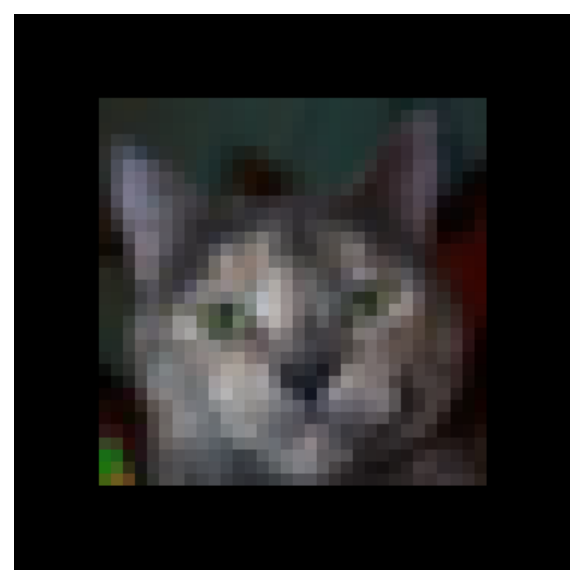}
        \includegraphics[width=0.95\linewidth]{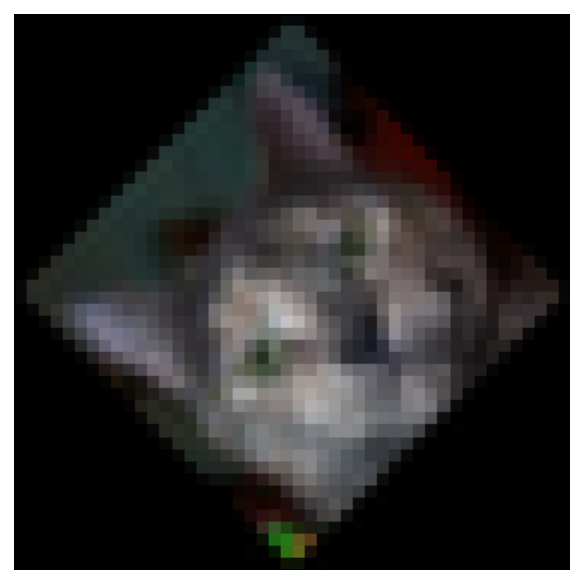}
    \end{minipage}
    \hfill
    \begin{minipage}{0.48\linewidth}
        \centering
        \includegraphics[width=0.95\linewidth]{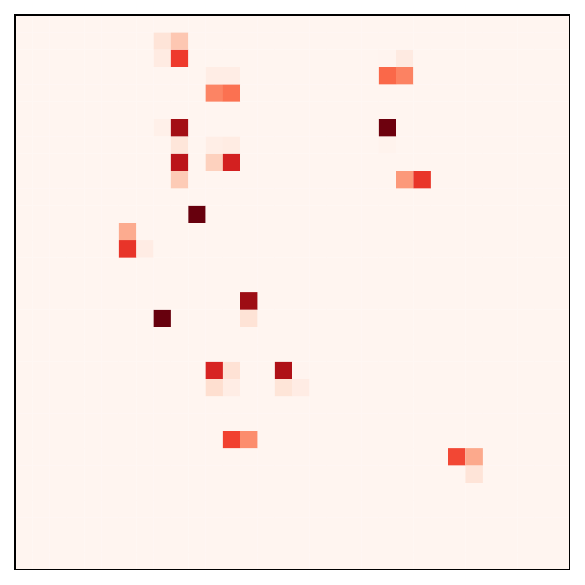}
        \includegraphics[width=0.95\linewidth]{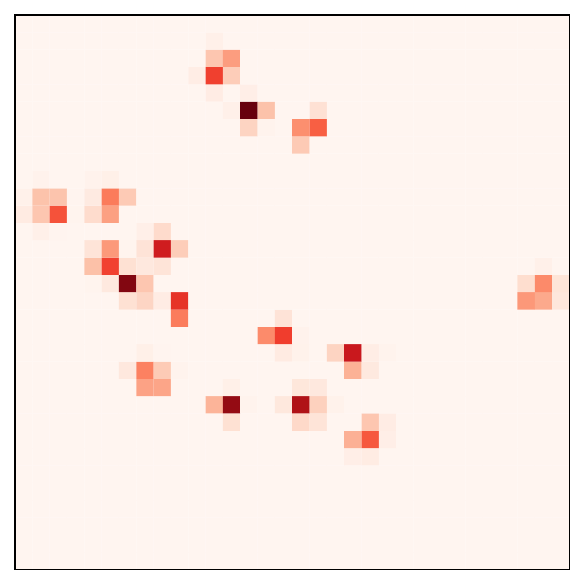}
    \end{minipage}
    \caption{ Demonstration of the significance of rotation-induced artifacts on less sparse (left) and more sparse (right) images by comparing original (top) and rotated images (bottom).  Visually, the image on the left (CIFAR-10\cite{krizhevsky:2009learning}) appears relatively unchanged after the rotation. The artifacts in the image on the right are far more prominent, and so might have relatively more influence on any learned strategy.}
    \label{fig:cifar_example}
\end{figure}

Certain tasks may be relatively more resilient to these artifacts than others.
For example, these effects will be less prominent for datasets containing high resolution images with pixels of uniform shape, where an approximate rotation will closely resemble the true rotation.
These artifacts may also be less impactful for non-sparse images, where the exact value of a given pixel could be less vital in observing macroscopic structure, as illustrated in Fig~\ref{fig:cifar_example}.
However, for calorimeter images, which may be low resolution, use non-uniform pixels, or be sparse, such artifacts are often significantly detrimental.
In pixelated calorimeter images, an artifact-free augmented dataset cannot not be created from the post-detector data alone; a network trained on the post-detector augmentations is learning the wrong symmetry. The true symmetry, before being broken, is not demonstrated in these augmentations~\cite{DBLP:journals/corr/abs-2002-08791}. 

We propose a novel data-augmentation technique which allows a network to learn a broken symmetry, assisting the learning process and increasing data efficiency. 
In traditional data augmentation, examples are synthesized directly from the post-detection training data, resulting in non-physical artifacts such that the augmented examples do not reflect the desired symmetry, making it challenging for the network to infer their relationship.
 Perhaps more importantly, the post-detector augmented data do not represent the expected detector patterns under the true detector transformation, such that a network which attempts to infer the symmetry is learning the wrong thing.  We introduce two modifications to the traditional approach: pre-detector augmentation and encouraged invariance.  If the data are generated with a simulator,  as is the case with calorimeter images, pre-detector augmentation applies the transformation before the symmetry is broken, rather than after; this is also referred to as re-simulation~\cite{Chirkin:2013avz}.  The post-detector data are still not closed under the transformation, but now accurately represent the set of expected detector signatures under the true symmetry.  Our second modification is to the loss function, penalizing the network for violating invariance across the pre-detector augmented examples of a given image during training. This effectively provides  crucial missing information, indicating to the network which set of images are expected to produce the same output. In this way the symmetry that was initially hidden by the detector is exposed during training, and the network might learn to classify in an approximately invariant manner. 
If the simulation used to create the training data represents the physical process well, then a network trained using this technique might be fine tuned and applied to real world data, though this step is not explored in this study. 
We explore the effectiveness of this technique on a simplified toy dataset designed to have similar properties to calorimeter images, which is computationally cheaper to produce than using a full simulation pipeline. 
In order to probe the relative data efficiency of each method, the performance of training with a post-detection augmented dataset is compared to that of a pre-detection augmented dataset, as well as to a network where output invariance is encouraged, across a range of training set sizes.

The organization of this paper is  as follows. 
Sec.~\ref{sec:dataset} provides the details of the dataset used, summarizing how it is generated and the structure of the resulting data. 
Sec.~\ref{sec:methods} covers the proposed method for encouraging model output invariance, as well as model implementation and evaluation.
Sec.~\ref{sec:results} presents numerical results along with their discussion.
Finally, Sec.~\ref{sec:conclusions} ends with conclusions, summarizing findings and future outlook.

\section{Dataset}
\label{sec:dataset}

To reduce the computational and time costs associated with generating jet images through a full simulation pipeline, we use a simplified toy dataset to evaluate the viability of the proposed methods. The structure of the signal and background are inspired by jet substructure tasks (e.g. ~\cite{Lu:2022cxg}), but not intended to be physically realistic.

This dataset is generated using Python v$3.10.11$\cite{Rossum:10.5555/1593511} and Numpy v$1.22.3$\cite{harris:2020array}.
A given toy example, referred to as an ``event'', is composed of a list of simulated energy deposits, or intensity values with associated 2D spatial coordinates.
In total, $3000$ events consisting of $16$ deposits each are generated, with half belonging to a signal class sample, and the other half belonging to a background sample.
The intensity values are drawn from a uniform distribution of values between $0$ and $1$ for both classes.
For background events, the location of every deposit is drawn from a uniform distribution over a disk of radius $1$.
For signal events, initially only a single deposit is drawn from the uniform disk distribution.
Subsequent deposits are then drawn from a 2D Gaussian centered on this deposit, with equal variances of $0.3$ in each dimension. 
If the location drawn happens to be outside of the radius of the disk, it is redrawn until it is inside the disk.
This results in the signal events having an internal structure distinct from what is found in the background sample, in a spirit similar to the task of tagging jets with sub-structure~\cite{Collado:2020ehf}. To increase the complexity and realism of the problem, noise is added to the signal events by drawing a number of additional events from the uniform disk distribution, independent of the other deposits within the event. The noise deposits only differ in their spatial distribution, with the intensities values being drawn from the same distribution across all deposits.
This means that the network cannot learn to separate the noise based on this information, and the overall normalization remains similar between the two classes.

\begin{figure}
    \centering
    \includegraphics[width=0.95\linewidth]{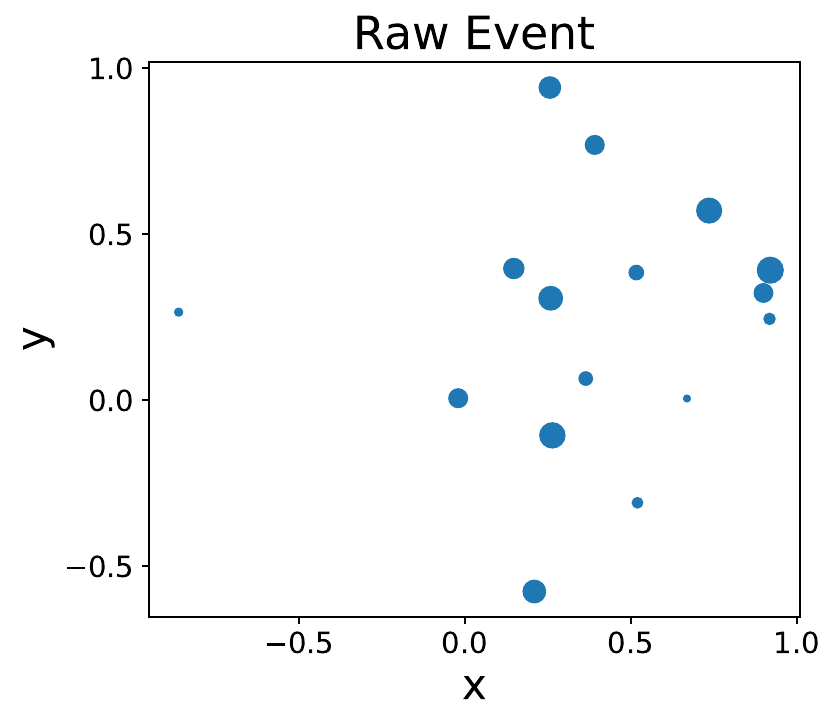}
    \includegraphics[width=0.95\linewidth]{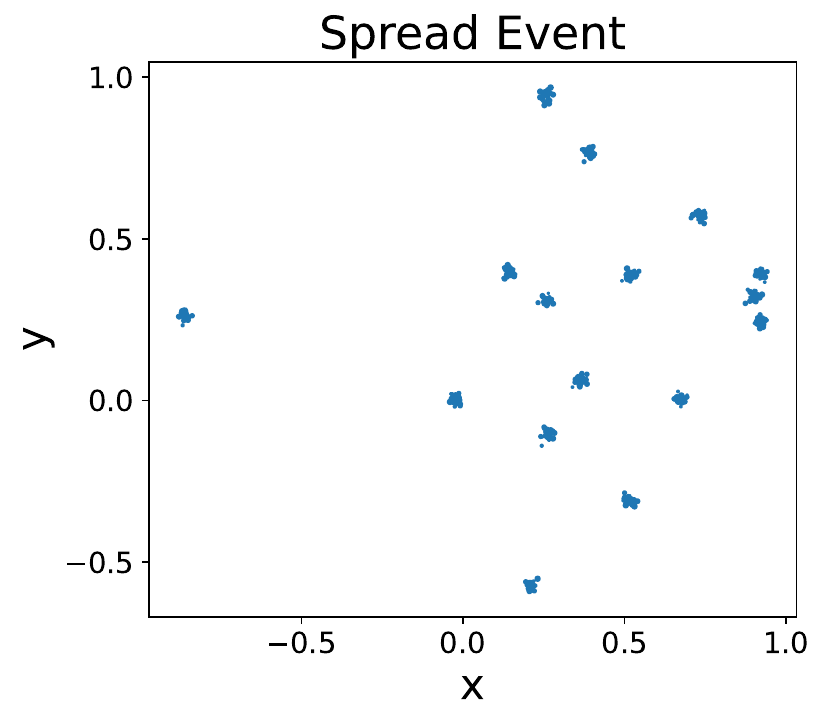}
    \caption{ A visualization of the dataset generation process. (Top) An example of an event before pixelization, where the size of each deposit is proportional to its energy. (Bottom) The same event, after the deposits have been distributed over a small area to simulate shower width effects which can lead to deposits over adjacent pixels. In the bottom pane the size of the deposits is arbitrary.}
    \label{fig:event_examples_raw}
\end{figure}

To mimic the physical extent of the shower created by an incident particle, each deposit is given a width by drawing $32$ points from Gaussian distributions, with equal variances of $\num{1e-4}$ in each dimension, centered on each of the chosen locations.
The intensity corresponding to a given deposit is then distributed evenly across these points. 
An example of an event generated before and after the spreading process may be seen in Fig.~\ref{fig:event_examples_raw}. 

\begin{figure}
    \centering
    \includegraphics[width=0.95\linewidth]{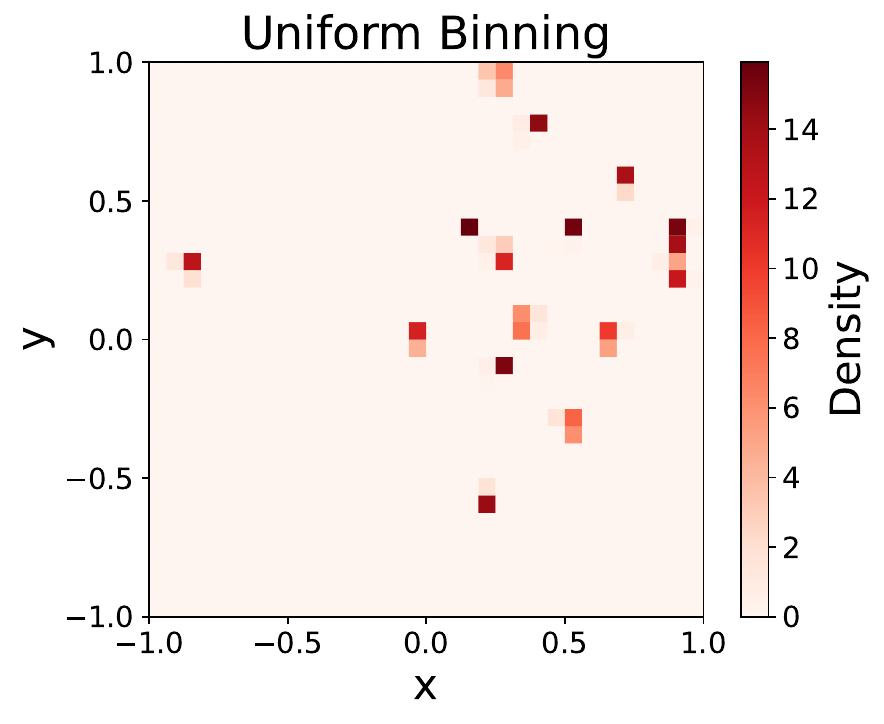}
    \includegraphics[width=0.95\linewidth]{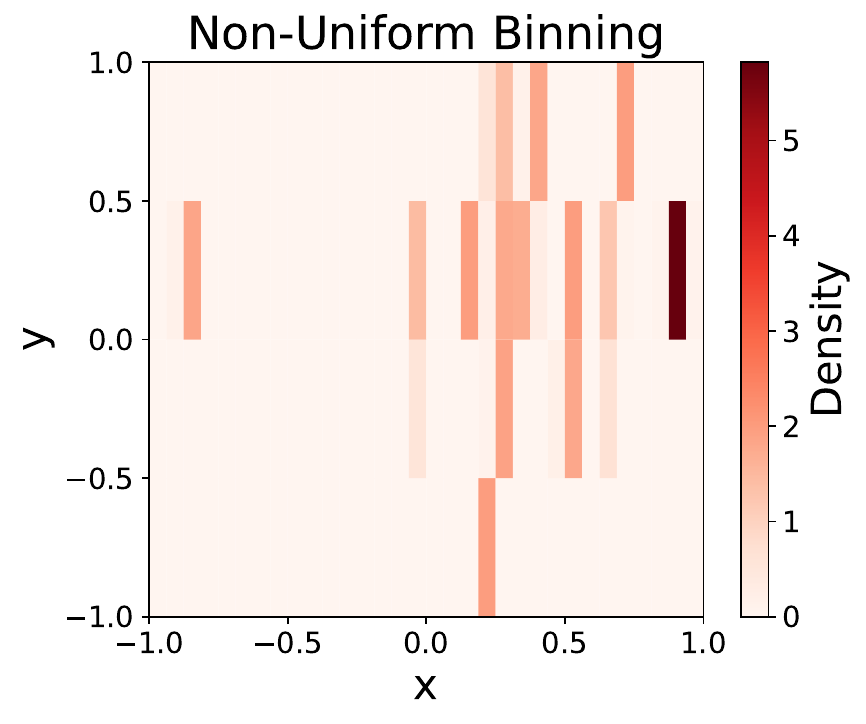}
    \caption{The event shown in Fig.~\ref{fig:event_examples_raw} with uniform binning (Top) and rectangular binning (Bottom) applied.}
    \label{fig:event_examples_binned}
\end{figure}

Pre-detector augmented images are created by applying rotations in $45^{\circ}$ increments to the events at this step, creating a set of $8$ copies for every event.
The intensities are then binned according to their position for a simplified detector response, creating the final pixelated event images.
Two variants of the pixelated events are created, with one using square binning on a $32\times32$ grid, and the other using rectangular binning on a $4\times32$ grid.
An example of a single event with each binning scheme is shown in Fig.~\ref{fig:event_examples_binned}.

\begin{figure}
    \centering
    \begin{minipage}{0.49\linewidth}
        \centering
        \includegraphics[width=1.15\linewidth]{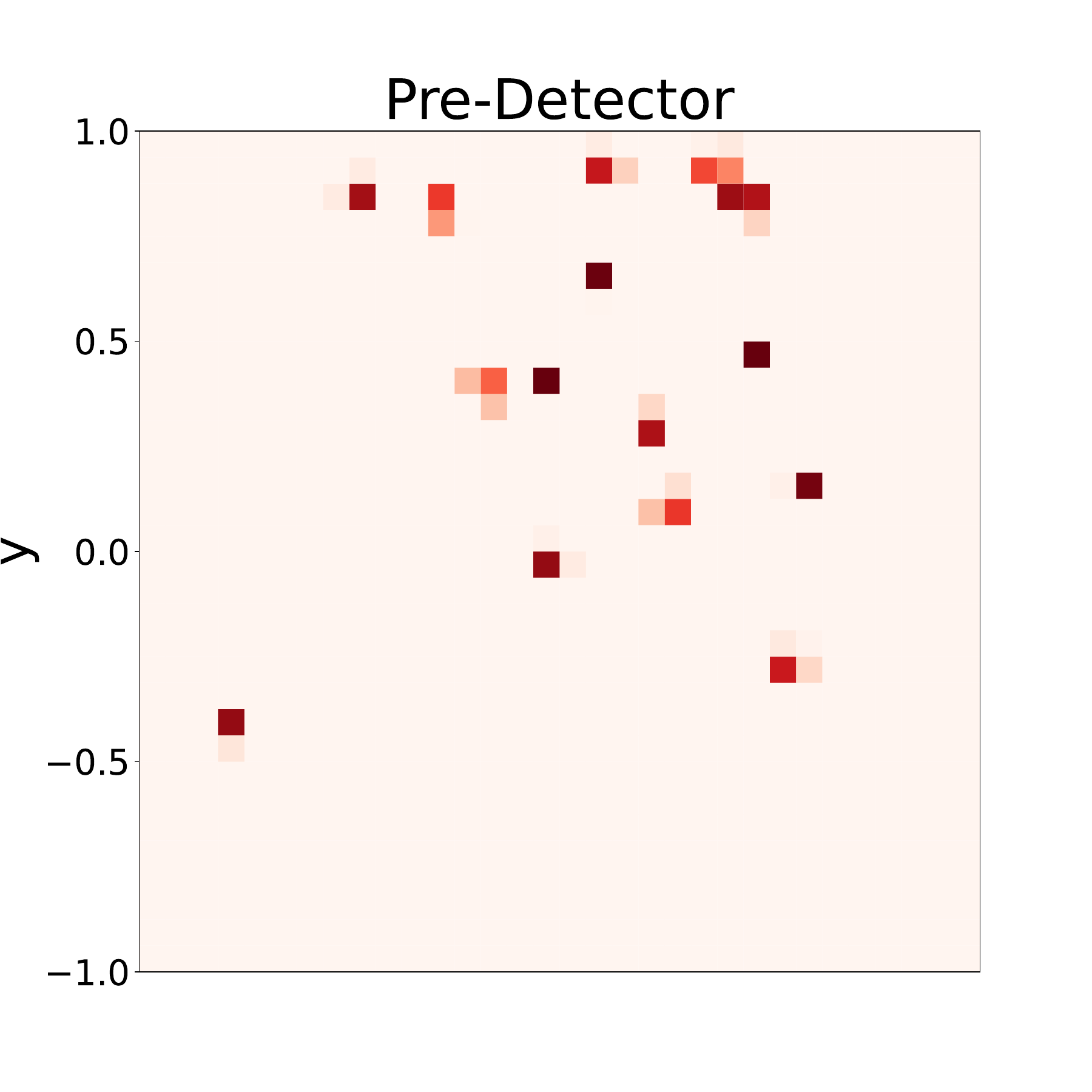}
        \includegraphics[width=1.15\linewidth]{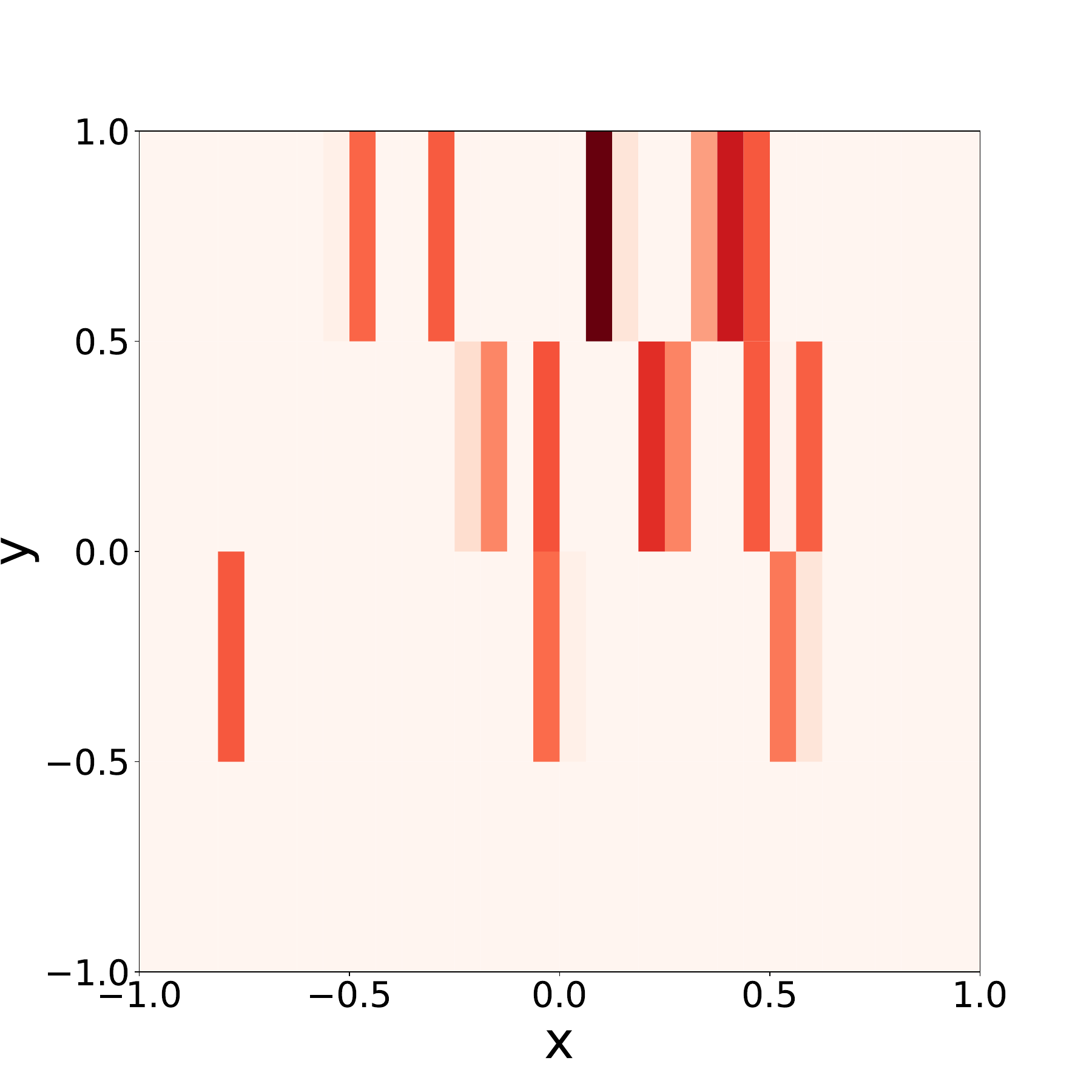}
    \end{minipage}
    \hfill
    \begin{minipage}{0.49\linewidth}
        \centering
        \includegraphics[width=1.15\linewidth]{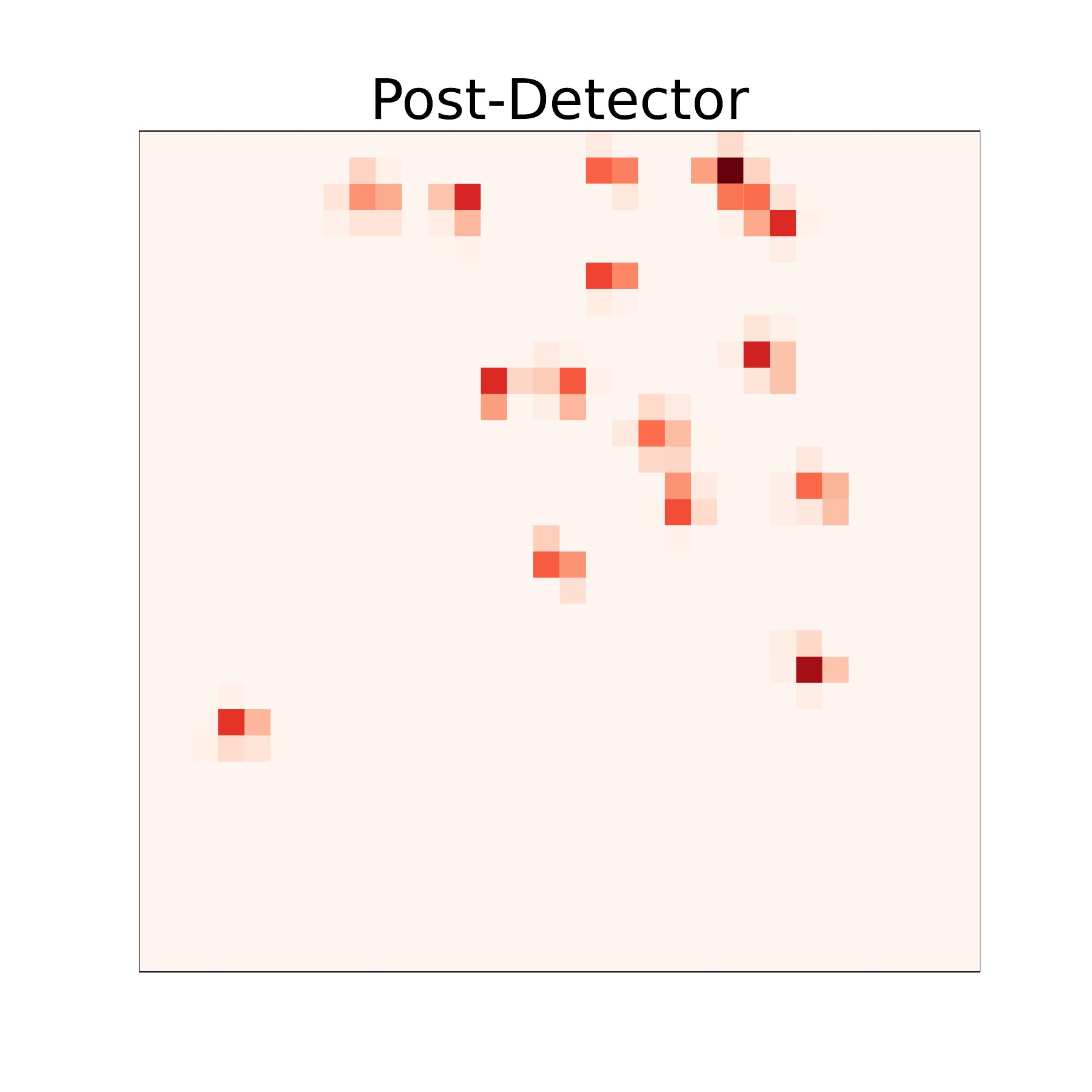}
        \includegraphics[width=1.15\linewidth]{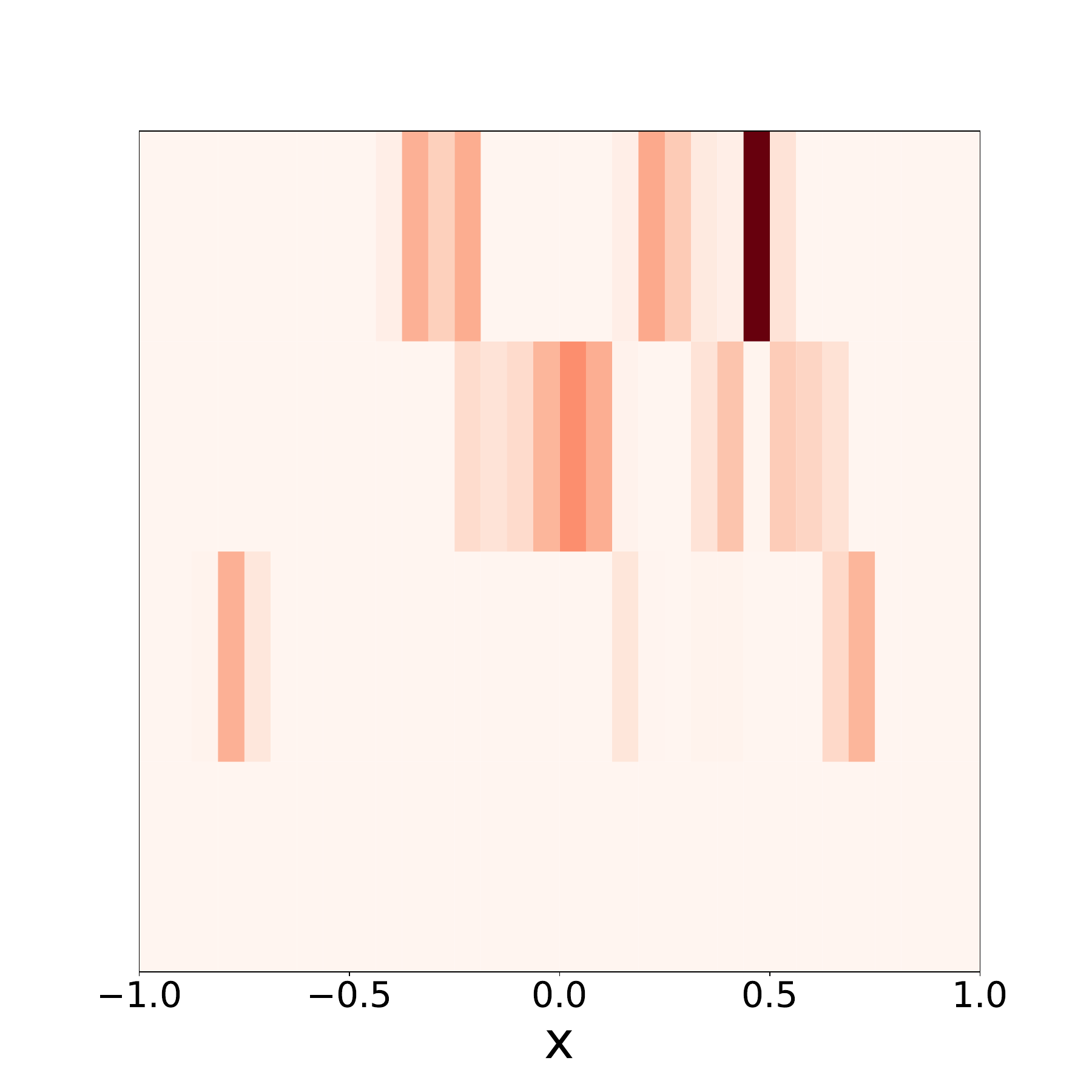}
    \end{minipage}
    \caption{The event shown in Fig.~\ref{fig:event_examples_raw} rotated at 45 degrees from their original orientation with uniform binning (Top) and rectangular binning (Bottom) applied. Applying rotations prior to binning (Left) avoids the interpolation artifacts which arise from applying rotations after binning (Right), where the images look relatively blurrier and more washed out.}
    \label{fig:event_examples_rotated}
\end{figure}

Rectangular pixel geometries are present in real detectors\cite{Gaudiello:201556}, and this non-uniformity could exaggerate the degree to which the effects of rotation are obscured by binning.
Traditionally, in computer vision tasks where augmentations are applied, it is done directly to pixelated images.
In our pipeline this would be equivalent to applying the rotation after the binning step. 
By applying rotations prior to the binning step, we are able to augment the dataset with synthetic examples which have been transformed exactly as expected, without introducing non-physical effects.
For comparison, we create another augmented dataset where augmentations are applied after the binning step using OpenCV v4.7.0\cite{opencv_library}, with the default bilinear interpolation method for rotating pixels.
It should be noted that other interpolation methods, as well as foregoing interpolation entirely in the case that re-binning is not required to produce valid network inputs, may result in different performance.
While the performance may change, none of these methods will be capable of recovering information lost through pixelization, and so will not match the output obtained by applying rotations pre-detection.
The differences between applying rotations before and after the detector step are demonstrated in Fig.~\ref{fig:event_examples_rotated}.
This leads to higher quality synthetic examples, as well as a way to expose a symmetry hidden by the binning to a neural network during training.
This may only be done in cases where the data is accessible prior to the step which results in pixelization, which could be the case for simulation but not for real world collider data.

\section{Encouraged Invariance}
\label{sec:methods}

We extend our novel learning method beyond the pre-detector data augmentation procedure outlined above, by  the use of a loss function which explicitly encourages a neural network to learn a classification invariant to augmentations applied prior to the binning. 
This is achieved by adding an additional component to the usual classification loss function, which penalizes differences in outputs across all of the augmented variants for a given event.
This takes the form shown in Eq.~\ref{eq:loss_function}, where $a$ and $b$ are scalar weight parameters, $L_{\text{cls}}$ is a typical loss function used for classification, and $L_{\text{inv}}$ is a loss function responsible for encouraging invariance.

\begin{equation}
\label{eq:loss_function}
L = aL_{\text{cls}} + bL_{\text{inv}}
\end{equation}

For all models presented here, binary cross-entropy is used for $L_{\text{cls}}$, and mean squared error is used for $L_{\text{inv}}$. 
Training is constrained to process all augmented copies for a given event within the same batch before a gradient update, producing a network output for each individual copy.
The standard deviation is computed across the outputs produced by augmented copies of a given event.
The outputs themselves are passed to $L_{\text{cls}}$, and the standard deviations are passed to $L_{\text{inv}}$. 
$L_{\text{inv}}$ penalizes for non-zero deviations, and in the case of mean squared error this is done by comparing the computed values to zero.
This effectively introduces more information to the training, by encouraging the network to treat all augmented versions of the jet identically, even if the post-detector data are not simply related under transformation in a way that would allow the network to infer it.

For models that are not trained with encouraged invariance, we effectively use $a=1$ and $b=0$ reducing the expression to the usual classification loss, and otherwise the weighting is tuned during hyperparameter optimization.
Fully Connected Networks (FCNs) and Particle Flow Networks\cite{Komiske:2018cqr} (PFNs), of the deep sets architecture\cite{zaheer2018deep}, are used to process the data, implemented and trained with Pytorch v$2.0.0$\cite{NEURIPS2019_9015}.
While the total number of pixels in an event is fixed, the number of non-zero pixels may differ between events. 
PFNs are a natural choice for this type of data, as they take a permutation invariant set of variable length inputs, and have been shown to yield good performance with collider data in previous studies\cite{Collado:2020ehf}.
The hyperparameters of each model are selected by performing $5$-fold cross-validation with a random search over a set of learning rates, batch sizes, layer sizes, and loss term weights with early stopping based on the validation loss.
Specifically, the learning rates are allowed to vary from $\num{1e-3}$ to $\num{1e-6}$ in steps of powers of $10$, batch sizes from $32$ to $1024$, layer sizes from $32$ to $512$ in steps of powers of $2$, and loss term weights are drawn from the set $\{0.01, 1, 10, 100, 300\}$.
A minimum change of $\num{1e-3}$ in the validation loss with a patience of $10$ epochs is used as the early stopping criteria.
This is done at the largest training size in order to obtain stable parameters in a computationally efficient manner.
Performance is then measured by constructing the standard Receiver Operating Characteristic (ROC) and evaluating the ROC AUC and signal efficiency at a fixed background efficiency of $50\%$, over $100$ event ensembles using the optimized hyperparameters.
Error due to statistical uncertainty is estimated from these ensembles to $1\sigma$.
Both architectures are evaluated in this way for the uniform and non-uniform image binning schemes across a range of training set sizes.
Each model is evaluated with augmentations applied prior to and after binning, and both augmentation schemes are evaluated with and without invariance encouraged.

\section{Results}
\label{sec:results}

\begin{figure*}
    \centering
    \includegraphics[width=0.45\linewidth]{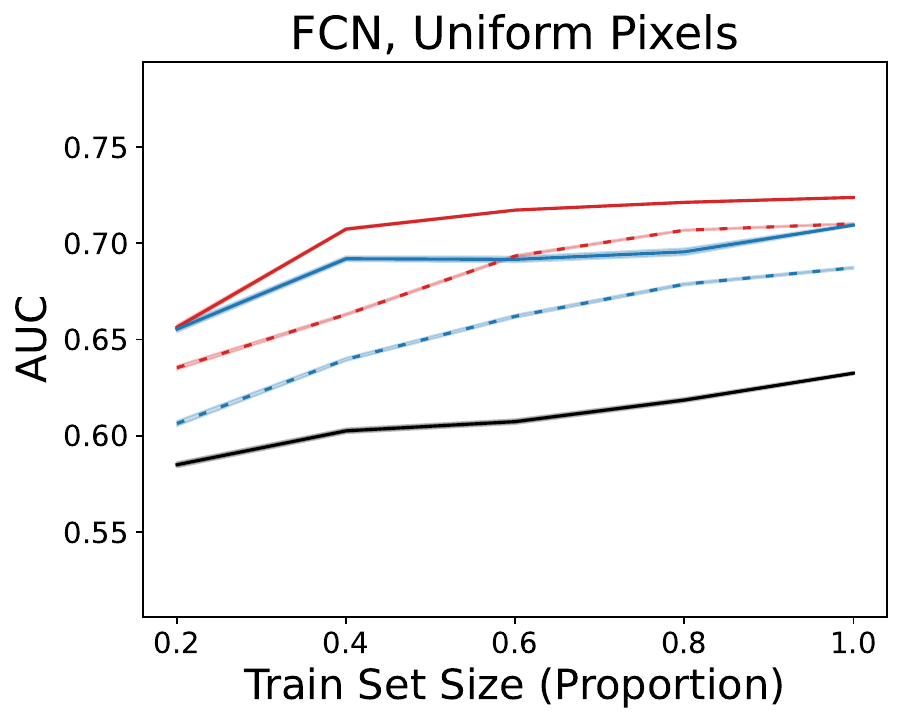}
    \includegraphics[width=0.45\linewidth]{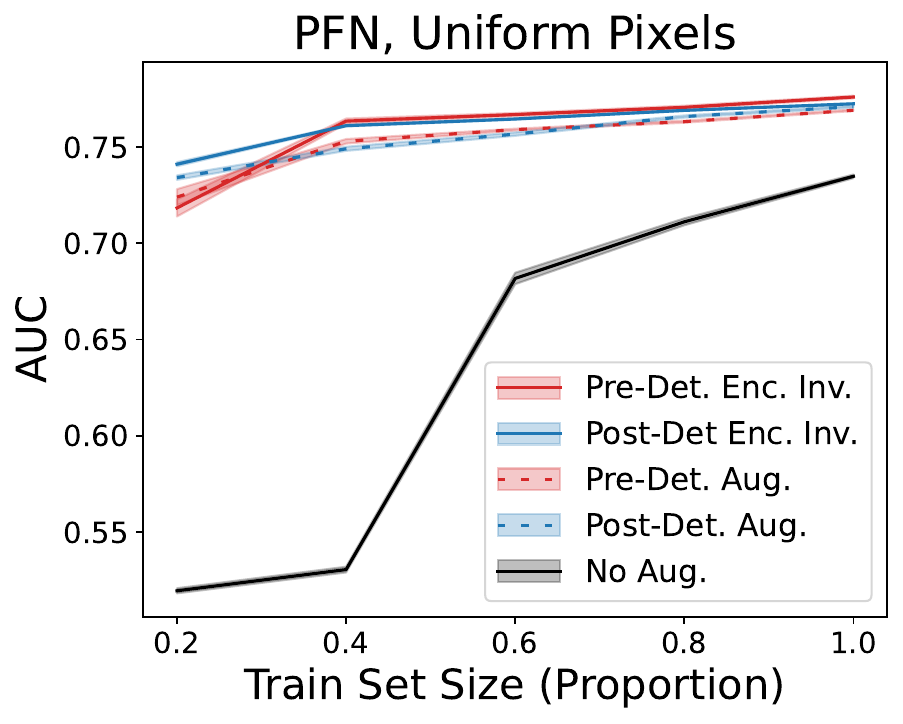}
      \includegraphics[width=0.45\linewidth]{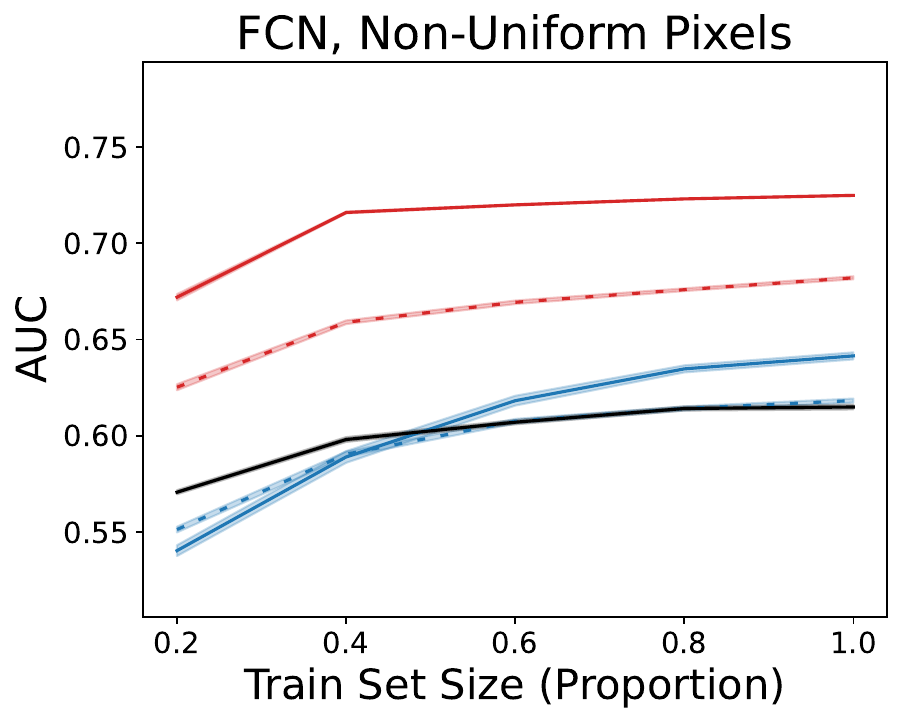}
    \includegraphics[width=0.45\linewidth]{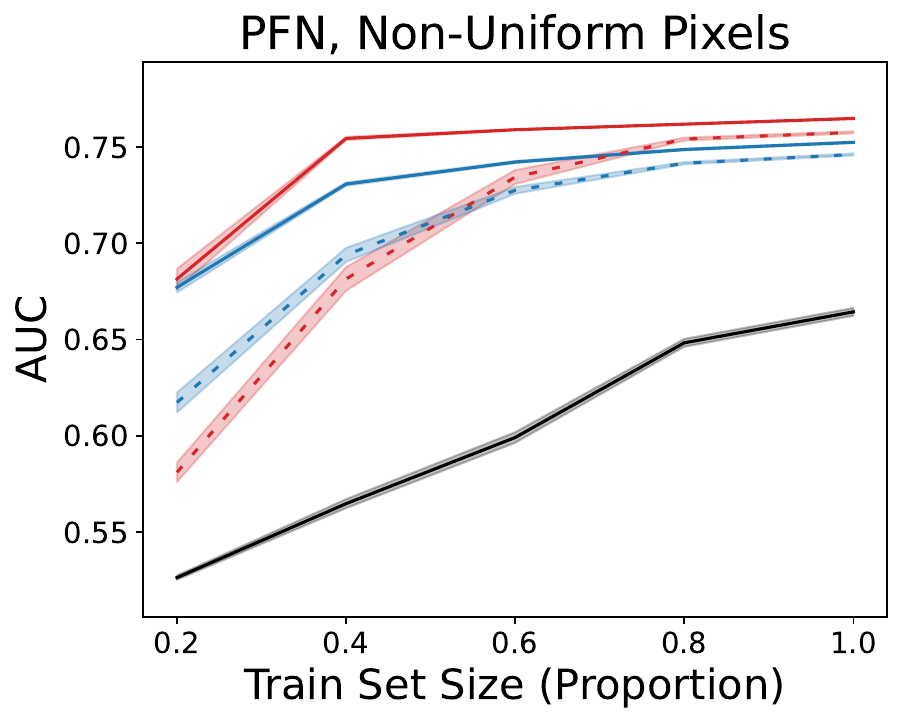}
    \caption{ROC AUC performance of FCNs (left) and PFNs (right) trained on uniformly binned data (top) or non-uniformly binned data (bottom) as a function of training set size. Though results depend on the nature of the task and the structure of the network, pre-detector augmentation and resimulation typically improves the learning rate, and encouraged invariance provides a further boost in learning. Colored bands represent the statistical variation, corresponding to one standard deviation ($1\sigma$)\, across 100 bootstrap ensembles.}
    \label{fig:train_size_scan_auc}
\end{figure*}

Learning strategies which take advantage of symmetries in the data can provide efficiencies in training, reaching a performance plateau with smaller training samples.  
We explore the relative power of several methods by evaluating the AUC with respect to the proportion of the full dataset used to train. 
Figure~\ref{fig:train_size_scan_auc} shows the dependence of the ROC AUC for FCNs and PFNs on the training set proportion for uniform pixels or non-uniform pixels. 
Additional figures and results for signal efficiency at $50\%$ background efficiency are included in the appendix.
For simplicity, we will focus on the AUC performance in the discussion that follows, as the trends found in the signal efficiency scans are very similar.

Scans are performed across a range of training set sizes, from $480$ to $3000$ unique events, not counting the $8$ synthetic variants of each image that are included in the data augmentation approaches.
We observe that in all cases the performance increases as the training set size increases, and that not using any augmentation yields the lowest performance, as might be expected.

In the case of uniform pixelization, the FCNs show a clear trend across the various methods.
At every training set size, there is a gain from one method to the next, with post-detection augmentations giving the smallest gain, followed by pre-detection augmentations, and then by encouraged invariance with the largest gain.
At every training set size, every method using augmentation yields an improvement over no augmentation.
Using examples augmented pre-detector results in a gain over post-detector augmentations, and encouraging invariance results in better performance than supplying augmented copies alone in both cases.
The PFN outperforms the FCN, which may not be surprising, as it is a more complex network that is better suited for this dataset.
Further, performance quickly converges for both augmentation schemes as well as encouraging invariance.
It appears that for uniformly pixelated data, where an approximate post-detection rotation leads to something close to a true pre-detection rotation, coupling any of the tested training methods with a more powerful network overcomes much of the challenge presented by the broken symmetry.

When non-uniform pixelization is used, differences between methods are apparent for both types of networks.
The FCN shows no clear gain in performance when using post-detection augmentations, but does benefit from pre-detection augmentations, seeing the largest improvement from using pre-detection augmentations with encouraged invariance. 
Interestingly, the PFN not only demonstrates a gain from using the post-detection augmented data, but shows a comparable gain from the pre-detection augmentations.
This suggests that there is useful information in the post-detection augmentations, but that it is harder for the FCN to take advantage of it, than it is for the PFN.
Unlike with uniform pixelization, the performance of the PFN does not completely converge across every method, but it does approach convergence more rapidly than the FCN.
As the training set size increases, the augmentation methods begin to converge to similar performance, with encouraging invariance on the pre-detector augmented dataset showing the highest gain.

These findings suggest that the differences in these methods are more apparent when training data is limited and when the symmetry breaking process is more exaggerated.
As including pre-detection augmented image variants in the training set, and encouraging invariance, consistently yield performance boosts for the FCN, these techniques may be especially beneficial in a computationally limited setting where the use of a simpler network is favored, even in the case that the symmetry of interest is only lightly obscured.
Notably, the performance gains obtained through the use of the pre-detection augmentations suggests that synthesizing new examples this way does indeed lead to higher quality augmented copies than applying the transformations post-detection.
It is of additional significance that across both pixelization schemes, encouraging invariance while using data augmented pre-detector almost always yields the best performance. 
This shows that training with explicit symmetry awareness meaningfully enhances the training process, in a way that is achievable even when the symmetry is obscured in the final data, to the point of not being exactly recoverable due to loss of information.
While these methods do require access to a way to apply transformations during data generation, it is not necessary to attempt to understand the specific details of how the transformation ends up being represented in the final data, which in itself may present an advantage.
The AUC values for the smallest and largest training set sizes are tabulated in Table~\ref{table:auc_summary}.

\begin{table}[h!]
    \centering
\begin{tabular}{llccccc}
\hline\hline
&  & \multicolumn{2}{c}{Uniform bins} & \ \ \ &\multicolumn{2}{c}{Non-uniform bins}\\ \cline{3-4} \cline{6-7}
& & small   & large  & & small   & large \\

Arch.& Augm. & set   & set  & & set   & set \\
\hline
   FCN & None & 0.585(1) & 0.632(1) && 0.571(1) & 0.615(1) \\
   & post-det. aug. & 0.606(2) & 0.687(1) && 0.551(1) & 0.618(1)  \\
   & pre-det. aug. & 0.635(1) & 0.710(1) && 0.625(2) & 0.682(1)\\
   & post-det. inv. & 0.655(2) & 0.709(1)&& 0.540(3) & 0.642(2)\\
   & pre-det. inv. & 0.656(1) & 0.724(1)&& 0.672(2) & 0.725(1)\\
   \hline
   PFN & None & 0.519(1) & 0.735(1) && 0.526(1) & 0.664(2)\\
   & post-det. aug. & 0.734(1) & 0.771(1) && 0.617(5) & 0.746(1)\\
   & pre-det. aug. & 0.724(4) & 0.770(1) && 0.581(5) & 0.758(1)\\
   & post-det. inv. & 0.741(1) & 0.772(1)&& 0.677(3) & 0.752(1)\\
   & pre-det. inv. & 0.718(4) & 0.776(1) && 0.681(5) & 0.765(1)\\
   \hline\hline
\end{tabular}
    \caption{The ROC AUC performance for models with various augmentation strategies described in the text, trained and evaluated on events with uniform or non-uniform pixelization, shown for the smallest and largest training set sizes tested.}
\label{table:auc_summary}
\end{table}

\section{Conclusions}
\label{sec:conclusions}

We propose a method for training neural networks with greater data efficiency, in the case that a symmetry known to be present in the data at some point during its generation is broken in the representation that is ultimately observed.
By creating an augmented dataset where the relevant transformation is applied at a step in the generation process when the symmetry is fully represented, higher quality synthetic examples may be obtained.
This information can be further leveraged by explicitly encouraging invariance across augmented variants of a given example through the loss during training.

We successfully demonstrate this method on a toy problem designed to probe the viability of these techniques for use with collider data.
In the case that a rotational symmetry is obscured by a detector-like binning process, training on a dataset which uses the higher quality augmentations results in better performance, especially with simpler networks.
Further, encouraging invariance can allow for even more data efficient training, showing that a network may be trained in a symmetry aware way even if the symmetry is not perfectly represented in the observed data.
Our results indicate that the utility of this technique depends on factors such as the amount of data available, the degree to which a symmetry is hidden within the data, and the type of network used.

Further work is necessary to determine how well this method can be used with both simulated and real world collider data.
Since pre-detector intervention and resimulation cannot be applied directly to real world data, it must also be explored how well performance gains are preserved when transferring from models trained on simulation.
This study also does not explore the optimization of the number of augmented copies created.
Using more augmented copies may serve to close some of the performance gap between the models which use synthetic examples, and those which encourage invariance.
This effect would likely be dependent on the dataset, and based on the degree to which performance is reliant on it, it may be worth optimizing along with other hyperparameters.

More generally, our findings suggest that it may be possible to improve the learning efficiency in other scenarios in which a symmetry is only approximately realized, such as Lorentz symmetry~\cite{Bogatskiy:2020tje,Butter:2017cot}.

\section{Acknowledgements}
\label{sec:acknowledgments}

The authors thank Chase Shimmin, Alexis Romero, Jason Baretz, and Kevin Greif for useful discussions and inspiring words. DW is supported by The Department of Energy Office of Science. EW is supported by his cat.

\section*{Code and Data}

The code for this paper can be found at \url{https://github.com/Edwit4/learning_broken_symmetries}. The datasets will be provided upon reasonable request to the authors.

\bibliography{main}

\clearpage
\appendix
\section{APPENDIX}
\label{sec:appendix}
\begin{figure*}
    \centering
    \includegraphics[width=0.45\linewidth]{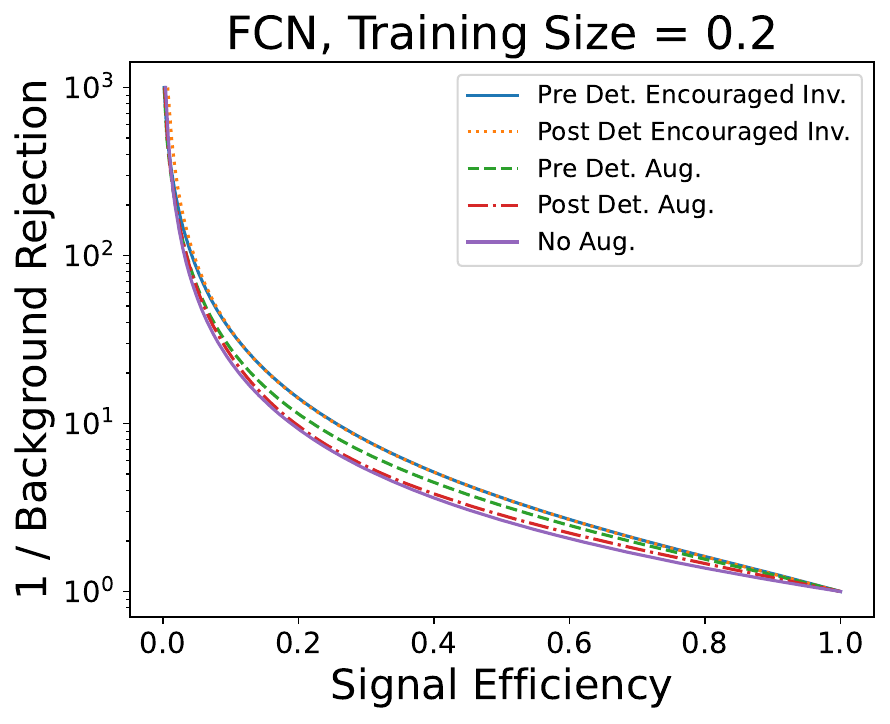}
    \includegraphics[width=0.45\linewidth]{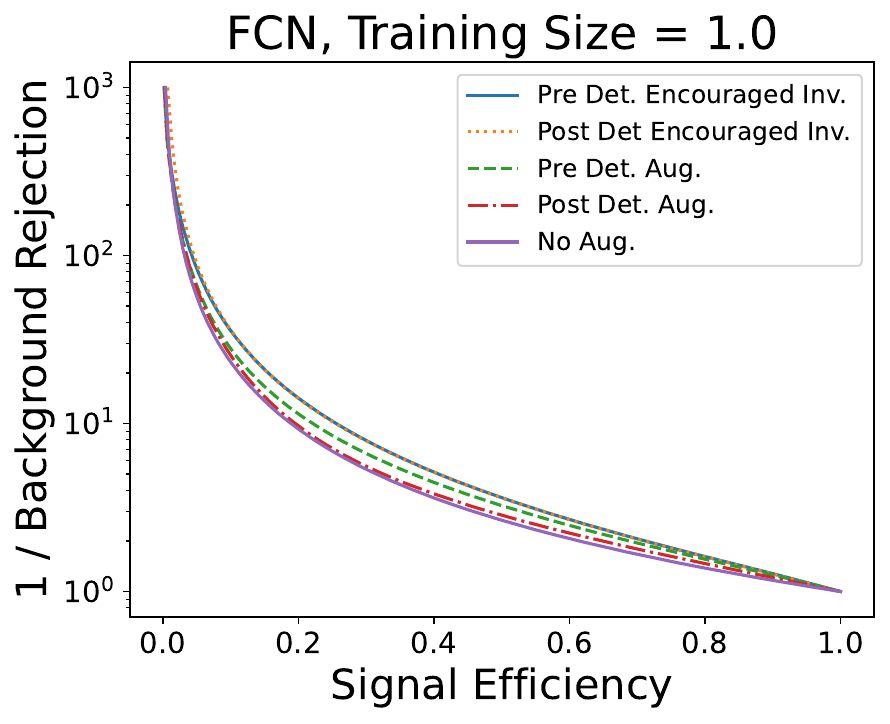}
      \includegraphics[width=0.45\linewidth]{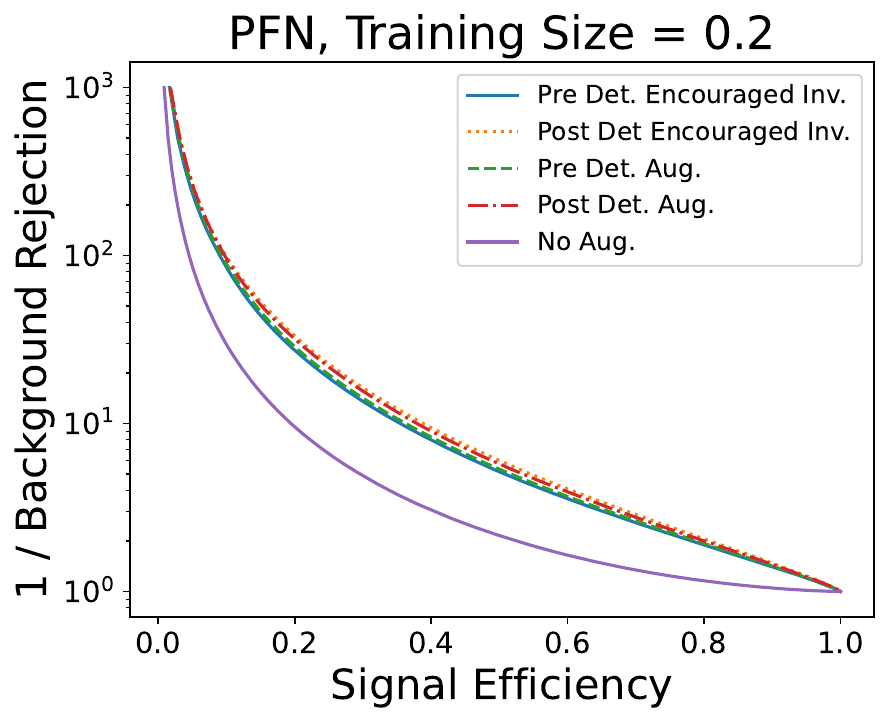}
    \includegraphics[width=0.45\linewidth]{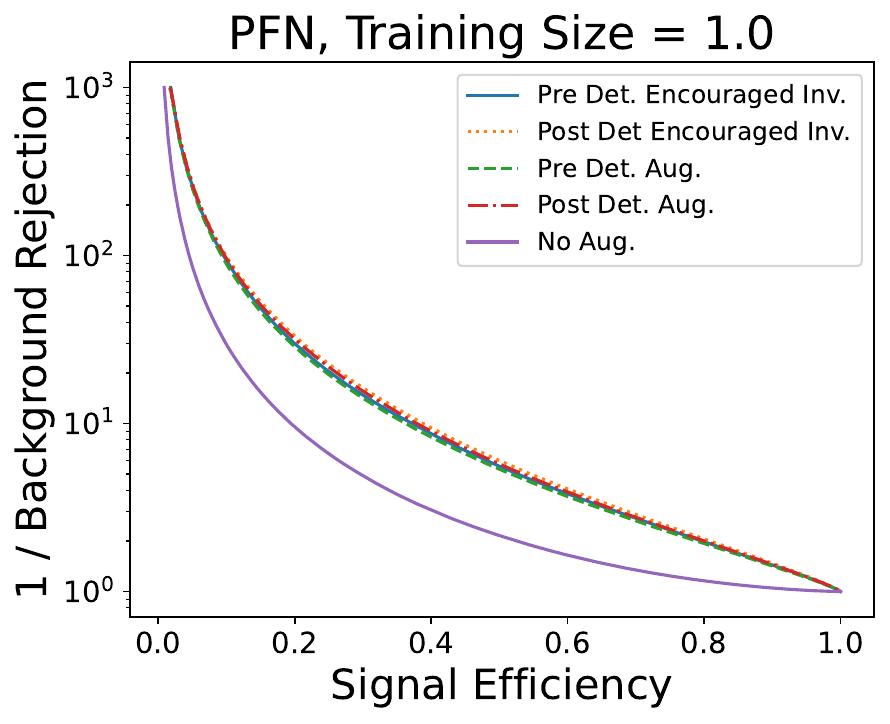}
    \caption{Performance, quantified through background rejection and signal efficiency, of FCNs (left) and PFNs (right) trained on a small training set size (top) or a large training set size (bottom) using uniformly binned data.}
    \label{fig:efficiency_rejection_uniform}
\end{figure*}

\begin{figure*}
    \centering
    \includegraphics[width=0.45\linewidth]{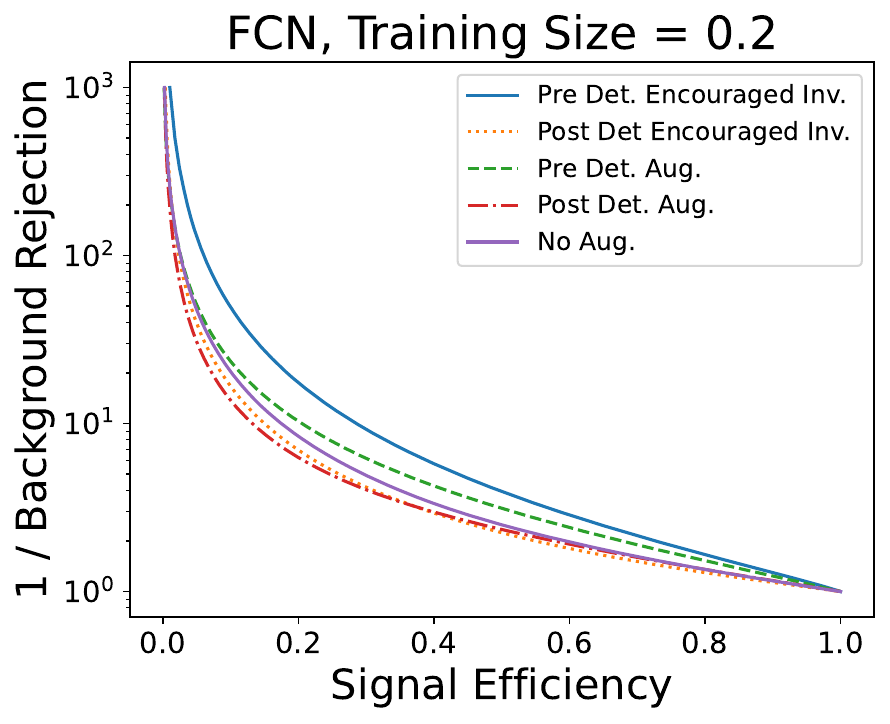}
    \includegraphics[width=0.45\linewidth]{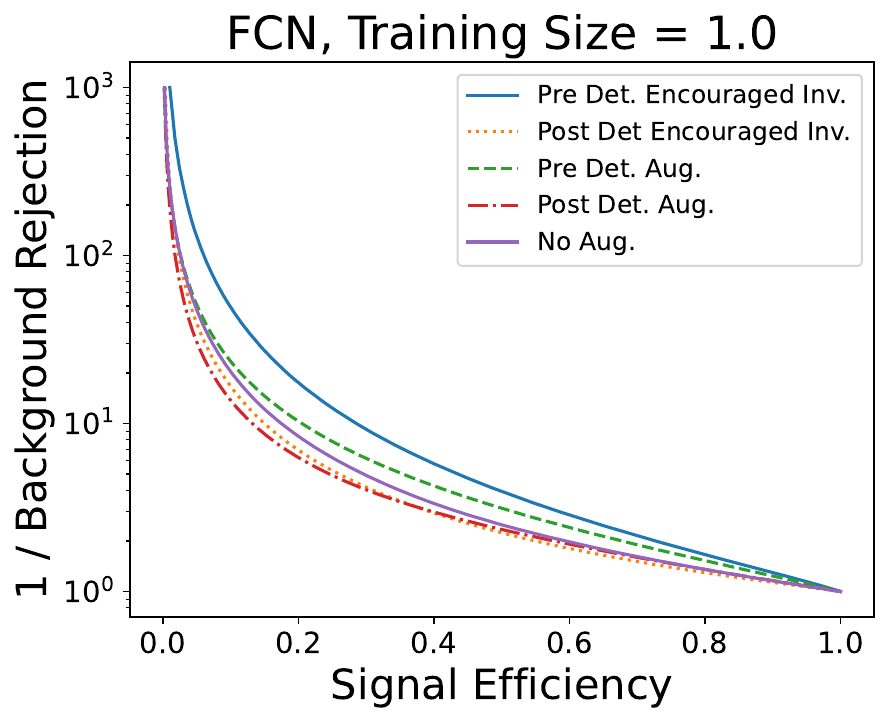}
      \includegraphics[width=0.45\linewidth]{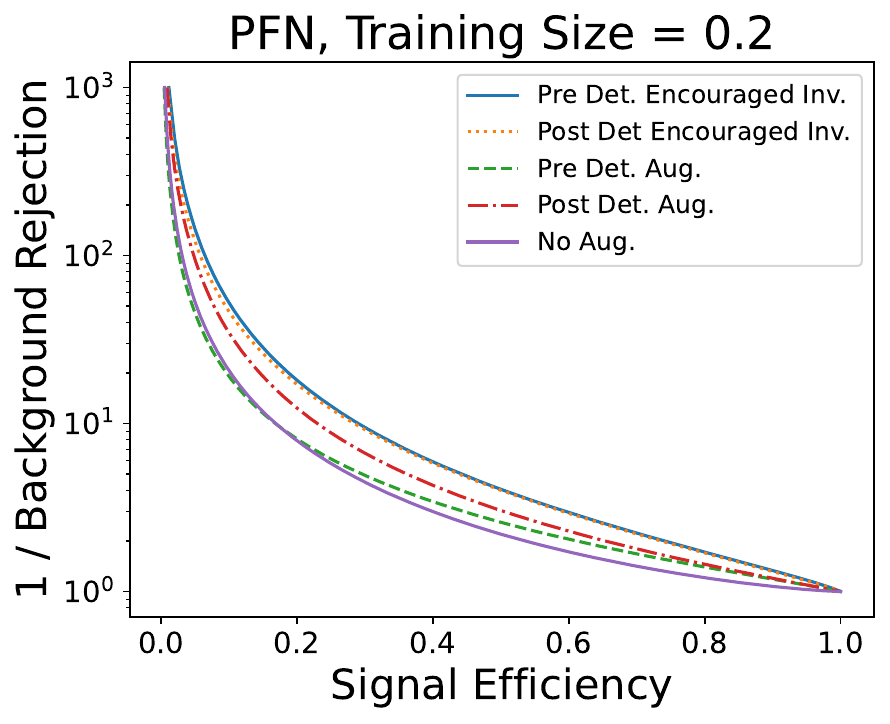}
    \includegraphics[width=0.45\linewidth]{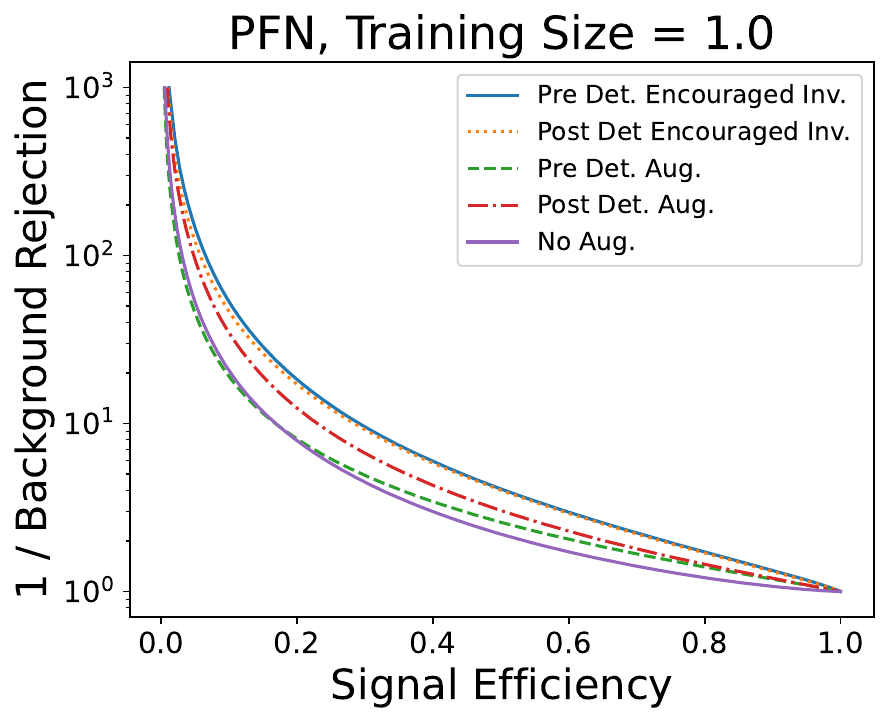}
    \caption{Performance, quantified through background rejection and signal efficiency, of FCNs (left) and PFNs (right) trained on a small training set size (top) or a large training set size (bottom) using non-uniformly binned data.}
    \label{fig:efficiency_rejection_rect}
\end{figure*}

Signal efficiencies are tabulated in Table~\ref{table:tpr_summary}. The results presented in these tables are illustrated in Figs.~\ref{fig:efficiency_rejection_uniform} and ~\ref{fig:efficiency_rejection_rect}.

\begin{figure*}
    \centering
    \includegraphics[width=0.45\linewidth]{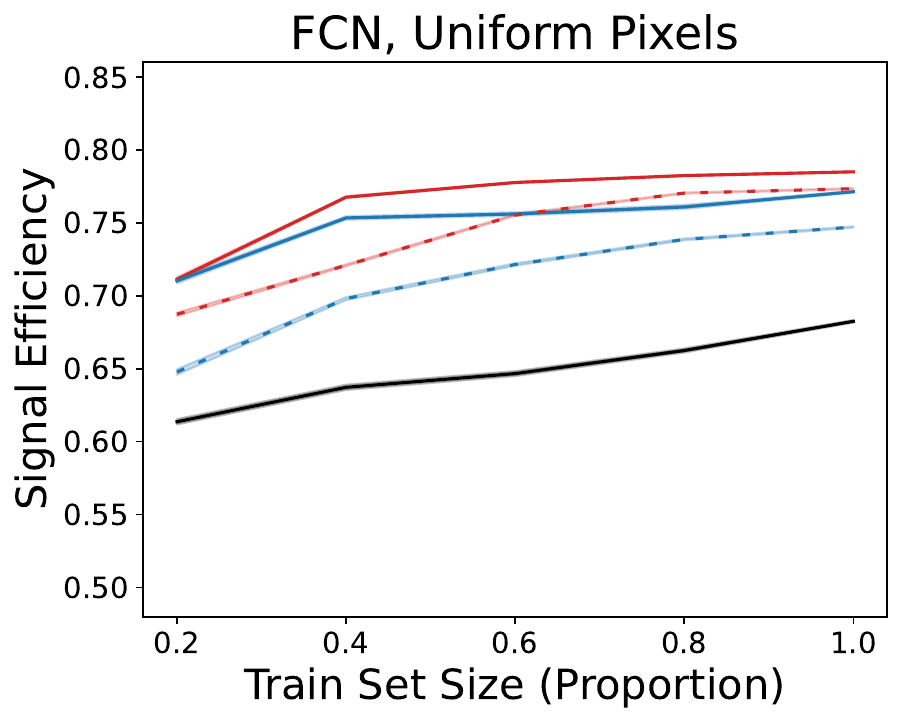}
    \includegraphics[width=0.45\linewidth]{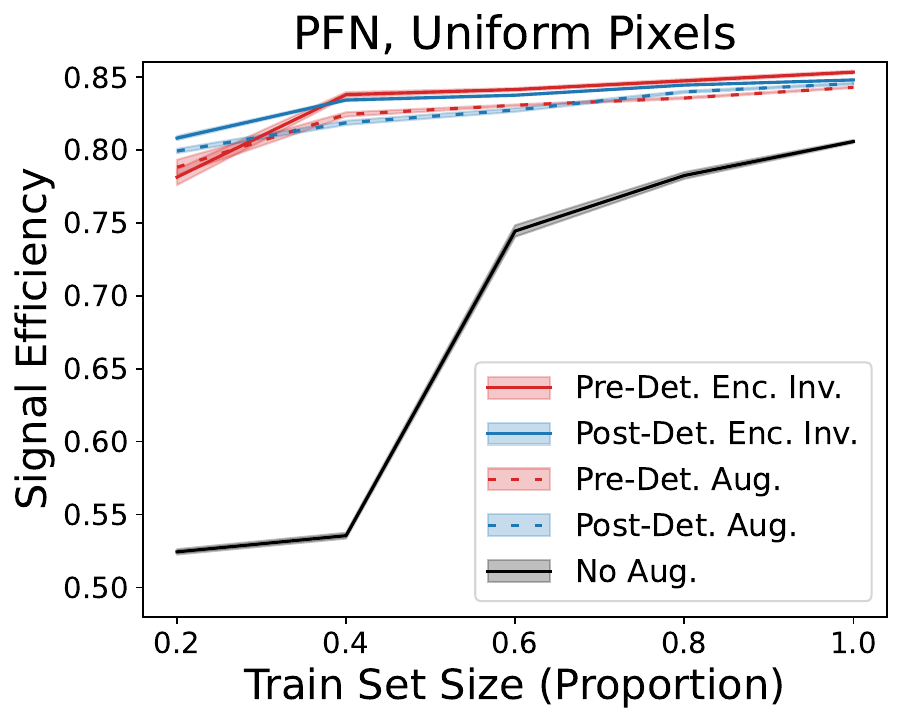}
      \includegraphics[width=0.45\linewidth]{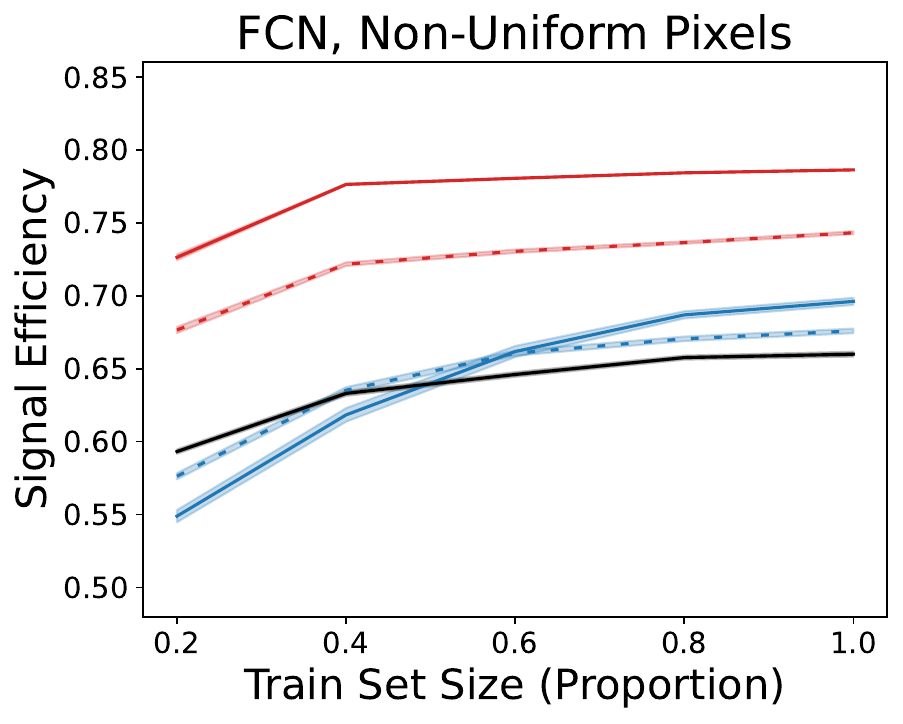}
    \includegraphics[width=0.45\linewidth]{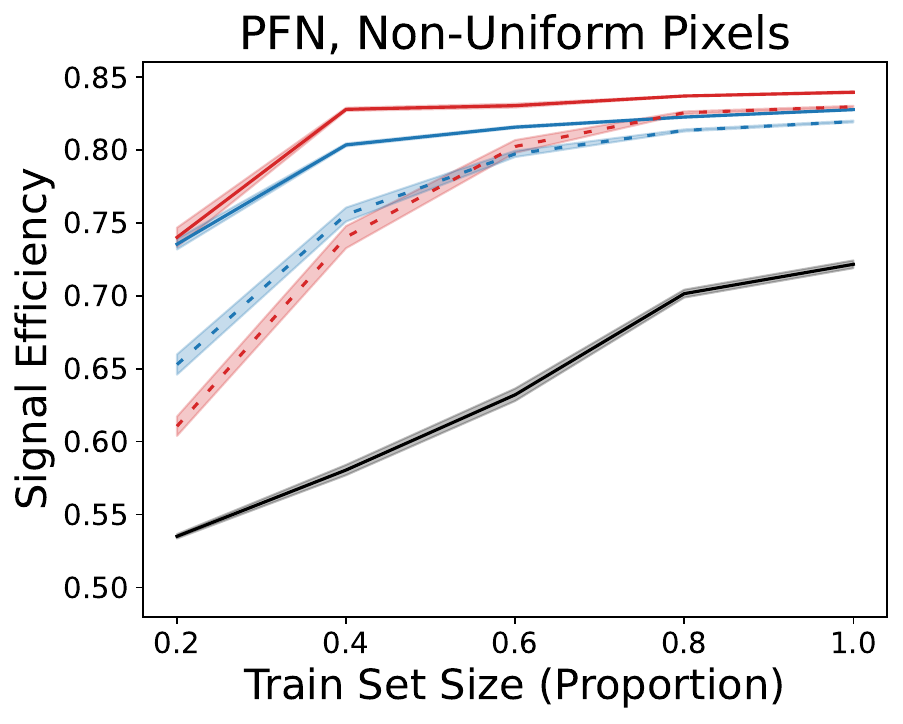}
    \caption{Signal efficiency, measured at a fixed background efficiency of 50\%, for FCNs (left) and PFNs (right) trained on uniformly binned data (top) or non-uniformly binned data (bottom) as a function of training set size. Colored bands represent the statistical variation, corresponding to one standard deviation ($1\sigma$)\, across 100 bootstrap ensembles.}
    \label{fig:train_size_scan_tpr}
\end{figure*}

The signal efficiencies at fixed background efficiency are depicted in Fig.~\ref{fig:train_size_scan_tpr} separated by network architecture and binning scheme.

\begin{figure*}
    \centering
    \includegraphics[width=0.45\linewidth]{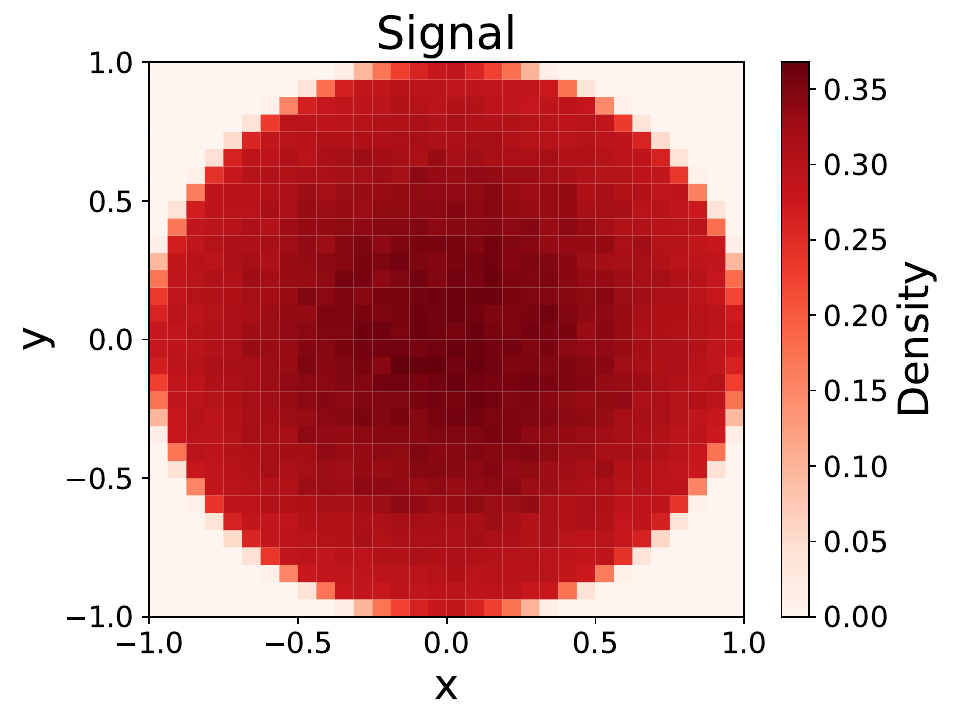}
    \includegraphics[width=0.45\linewidth]{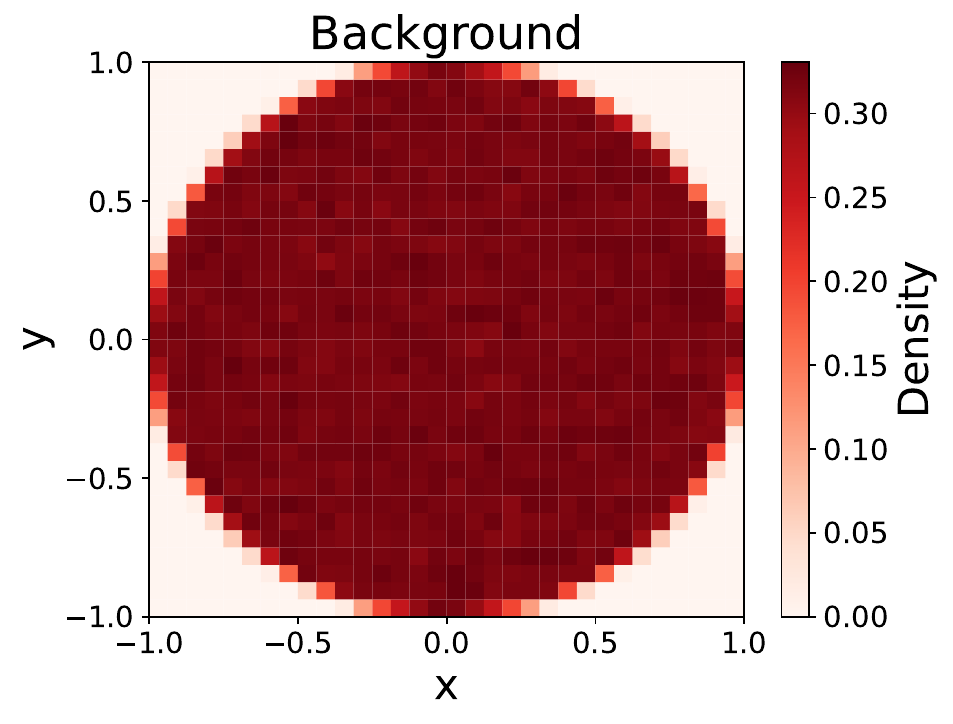}
      \includegraphics[width=0.45\linewidth]{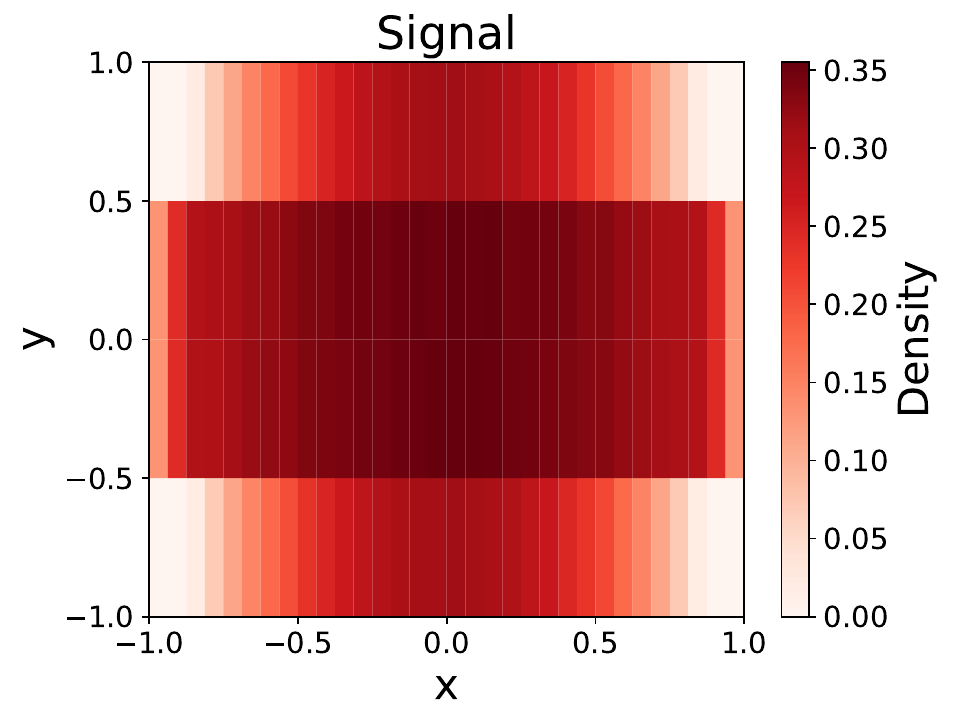}
    \includegraphics[width=0.45\linewidth]{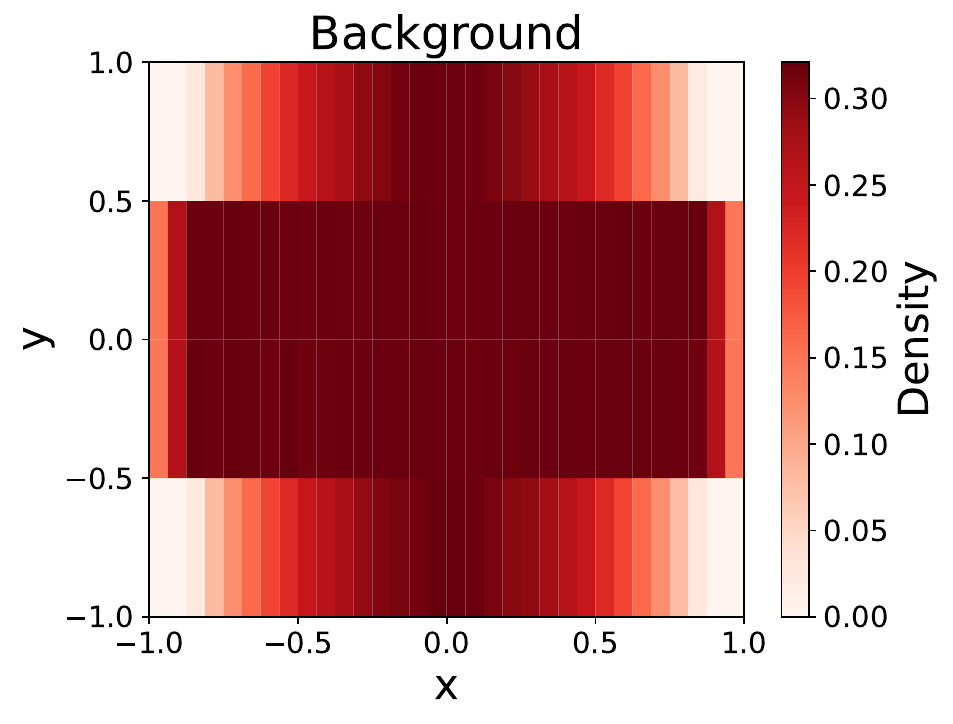}
    \caption{Average images of the signal (Left) and background samples (Right), with uniform pixelization (Top) and non-uniform pixelization (Bottom).}
    \label{fig:avg_images}
\end{figure*}

The averages of each sample may be seen in Fig.~\ref{fig:avg_images} with uniform binning and non-uniform binning.

\begin{table}[h!]
    \centering
\begin{tabular}{llccccc}
\hline\hline
&  & \multicolumn{2}{c}{Uniform bins} & \ \ \ &\multicolumn{2}{c}{Non-uniform bins}\\ \cline{3-4} \cline{6-7}
& & small   & large  & & small   & large \\

Arch.& Augm. & set   & set  & & set   & set \\
\hline
   FCN & None & 0.614(2) & 0.682(1) && 0.593(1) & 0.660(2) \\
   & post-det. aug. & 0.648(2) & 0.747(1) && 0.576(2) & 0.676(2) \\
   & pre-det. aug. & 0.687(1) & 0.773(1) && 0.677(2) & 0.743(1) \\
   & post-det. inv. & 0.710(2) & 0.771(1) && 0.549(4) & 0.696(2) \\
   & pre-det. inv. & 0.711(1) & 0.785(1) && 0.726(2) & 0.786(1) \\
   \hline
   PFN & None & 0.524(2) & 0.806(1) && 0.535(2) & 0.722(3) \\
   & post-det. aug. & 0.799(1) & 0.845(1) && 0.653(7) & 0.819(1) \\
   & pre-det. aug. & 0.788(5) & 0.843(1) && 0.611(7) & 0.830(1) \\
   & post-det. enc. & 0.808(1) & 0.848(1) && 0.735(4) & 0.828(1) \\
   & pre-det. enc. & 0.781(5) & 0.853(1) && 0.740(7) & 0.840(1) \\
   \hline\hline
\end{tabular}
    \caption{A summary of the signal efficiency at a fixed background efficiency of $50\%$, shown for the smallest and largest training set sizes tested and both pixelization schemes.}
\label{table:tpr_summary}
\end{table}

\end{document}